# Structural and Magnetic Properties of Nanosized Barium Hexaferrite Powders Obtained by Microemulsion Technique

Tatyana Koutzarova [1, a], Svetoslav Kolev [1, b], Kornely Grigorov [1, c], Chavdar Ghelev [1, d], Andrzej Zaleski [2, e], Robert E. Vandenberghe [3, f], Marcel Ausloos [4, g], Catherine Henrist [5, h], Rudi Cloots [5, i] and Ivan Nedkov [1, j]

[1] Institute of Electronics, Bulgarian Academy of Sciences, 1784 Sofia, Bulgaria

[2] Institute of Low Temperatures and Structural Research, Polish Academy of Sciences, 50-422 Wroclaw, Poland

[3] Department of Subatomic and Radiation Physics, University of Gent, 9000 Gent, Belgium

[4] SUPRATECS, Sart Tilman, B-4000 Liege, Belgium

[5] LSIC, Chemistry Department B6, University of Liege, Sart Tilman, B-4000 Liege, Belgium

[a]tanya@ie.bas.bg, [b]skolev@ie.bas.bg, [c]kgrigoro@abv.bg, [d]chghelev@ie.bas.bg, [e]a.zaleski@int.pan.wroc.pl, [f]robert.vandenberghe@ugent.be, [g]marcel.ausloos@ulg.ac.be, catherine.henrist@ulg.ac.be, [i]rcloots@ulg.ac.be, [j]nedkov@ie.bas.bg

**Abstract.** Thin hexagonal barium hexaferrite particles synthesized using the microemulsion technique were studied. A water-in-oil reverse microemulsion system with cetyltrimethylammonium bromide (CTAB) as a cationic surfactant, n-butanol as a co-surfactant, n-hexanol as a continuous oil phase, and an aqueous phase were used. The microstructural and magnetic properties were investigated. The particles obtained were mono-domain with average particle size 280 nm. The magnetic properties of the powder were investigated at 4.2 K and at room temperature. The saturation magnetization was 48.86 emu/g and the coercivity, $2.4 \times 10^5$ A/m at room temperature. The anisotropy field $H_a$ and magneto-crystalline anisotropy $K_1$ were $1.4 \times 10^6$ A/m and $2.37 \times 10^5$ J/m$^3$, respectively.

## Introduction

Barium hexaferrite particles are one of the most promising materials for high-density magnetic recording media due to their unique recording characteristics, namely, high coercivity, moderate magnetic moment, low or positive temperature coefficient of coercivity, and high chemical stability [1-3].

The *M*-type barium hexaferrite (BaFe$_{12}$O$_{19}$) is the hexaferrite family's best known compound. Its crystal structure is the so-called magnetoplumbite structure that can be described as a stacking sequence of the basic S (spinel) and R (hexagonal) blocks [4, 5]. The magnetic ion (Fe$^{3+}$) occupies five different interstitial positions in a ferrimagnetic order resulting in a net magnetic moment. Two of the possible 16 tetrahedral positions (4$f_1$) and four of the possible octahedral positions (2$a$) are occupied by Fe$^{3+}$ in the S block. Fe$^{3+}$ in the R block occupies octahedral sites in the octahedra shared by common faces (12$k$), in octahedra at the interface of adjacent blocks (4$f_2$), and trigonal bipyramidal sites (2$b$). The presence of magnetic Fe$^{3+}$ cations in these positions is responsible for the BaFe$_{12}$O$_{19}$ magnetic properties and for its magneto-crystalline anisotropy ($K_1 = 3.3 \times 10^5$ J/m$^3$) [6].

The physical properties of an inorganic microstructure are fundamentally related to the size, crystal structure and morphology, which can vary depending on the preparation route [7]. The traditional methods to prepare nanoparticles are rather complicated; they involve a number of different steps with multiple microstructural problems that may have a detrimental effect on the magnetic performance. This is one of the reasons why researchers keep on looking for new routes of synthesis and improvement of the known ones. The most commonly used among these "new routes" is synthesis of nano-sized powders by using "wet chemistry", often called the "chemical route". It is known that the composition, shape and size of the precursor particles used for high-temperature synthesis affects the microstructural characteristics of the material produced. Co-precipitation is one of the techniques used frequently for preparation of nanosized particles. The co-precipitation allows one to vary the average size of nanoparticles by adjusting the pH and the temperature of the aqueous media, but one has only limited control over the particles size distribution [8]. Recent investigations demonstrated the possibility to prepare homogeneous nanosized magnetic oxide powders by applying the microemulsion process [9, 10]. A microemulsion system consists of an oil phase, a surfactant phase and an aqueous phase. The reverse microemulsion system exhibits a dynamic structure of nanosized aqueous droplets which are in constant deformation, breakdown, and coalescence. Each of the aqueous droplets can act as a nanosized reactor for forming nanosized precipitate particles [7]. One of the advantages of this technique is the preparation of very uniform precursors' particles (< 10% variability) [8].

Since the condition of synthesis affect considerably the chemical, structural and physical properties, our attention was focused on investigating the microstructural and magnetic properties of $BaFe_{12}O_{19}$ powder synthesized by reverse microemulsion technique.

**Experiment**

A water-in-oil reverse microemulsion system with cetyltrimethylammonium bromide (CTAB), (24 wt.%) as a cationic surfactant, n-butanol (16 wt.%) as ca o-surfactant, n-hexanol (20 wt.%) as a continuous oil phase, and an aqueous solution (40 wt.%) was used. The metallic ions ($Ba^{2+}$ and $Fe^{3+}$) concentration in the aqueous phase was 0.44 M. The molar ratio of Ba to Fe was fixed at 1:10.

In the first step of the synthesis procedure, the co-precipitation occurred when the microemulsion containing an aqueous solution of $Ba(NO_3)_2$ and $FeCl_3$ was added to the microemulsion containing the precipitating agent NaOH. The amount of NaOH was set to a value that resulted in the final pH value after precipitation being 11. The precipitate obtained was separated in a centrifuge and was washed with water and solution of chloroform and methanol (50 v.% and 50 v.%) to remove the excess surfactant. The hydroxide precursor was dried and milled.

In the second step of the synthesis the powder obtained was heated at 580°C for 4 h. After grinding, the powder was finally calcined at 900°C for 5 h to ensure complete conversion of the precursors into $BaFe_{12}O_{19}$.

The barium hexaferrite powder was characterized using XRD analysis (TUR diffractometer with Bragg-Brentano geometry at room temperature using Cu-$K_\alpha$ radiation) and scanning electron microscopy (SEM, Philips ESEM XL30 FEG). The Mössbauer spectra were obtained by a conventional home-made spectrometer. A 50 mCi $^{57}$Co (Rh) source was used. The magnetic measurements were carried out at room temperature and at 4.2 K using a vibration sample magnetometer with a maximum magnetic field of 2.3 x $10^6$ A/m. The high magnetic field measurements (up to 1 x $10^7$ A/m) were performed on a homemade pulsed magnetometer [11]. The magnetic measurements were done on an unoriented random assembly of particles.

**Results and discussions**

The XRD spectrum of the synthesized $BaFe_{12}O_{19}$ powder is presented in Fig. 1. It shows the characteristic peaks corresponding to the barium hexaferrite structure. No other phases are detected. This confirms the complete conversion of the precursor powder into $BaFe_{12}O_{19}$. The lattice constants obtained from the XRD spectra are $a = 0.584$ nm and $c = 2.341$ nm.

Fig. 2 shows the morphology of the calcined powder. It exhibits a narrower grain-size distribution, with the average particle size being 280 nm. Most of the particles have an almost perfect hexagonal shape. It can be seen that the smallest particles with a size of about 250 nm have irregular shape because the process of forming the platelet's hexahedral shape typical for the $BaFe_{12}O_{19}$ has not been completed. The presence is also observed of spheroidal particles with a size of about 100 nm, their content in the sample being < 12%. The critical diameter ($D$) for single-domain barium hexaferrite particles is about 460 nm [12], so the particles of the sample are single domain. It is interesting to note that the particles prepared by this method are thin, their average thickness $t$ is 36 nm, and the average aspect ratio ($D/t$) is 7.7.

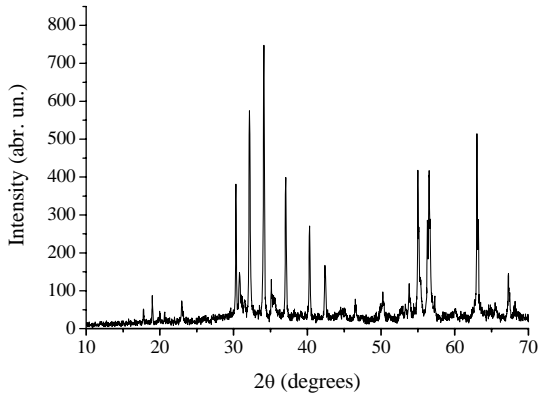
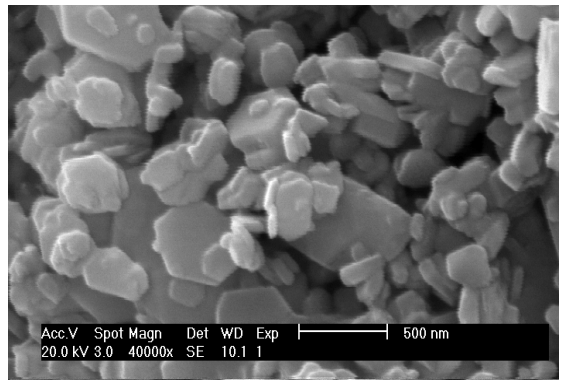

Fig. 1. X-ray diffraction pattern of $BaFe_{12}O_{19}$ powder.

Fig. 2. SEM image of $BaFe_{12}O_{19}$ powder.

The hysteresis loops of a sample at 4.2 K, room temperature and a maximum applied field of $2.3 \times 10^6$ A/m are shown in Fig. 3. The magnetic parameters, namely, the remanent magnetization ($M_r$) and coercivity field ($H_c$) obtained from the hysteretic curves, are listed in Table 1. The saturation magnetization values ($M_s$) were obtained from the magnetization curves in high magnetic fields up to $1 \times 10^7$ A/m and are presented on Fig. 4.

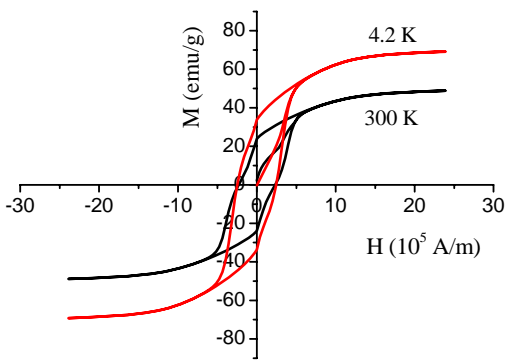
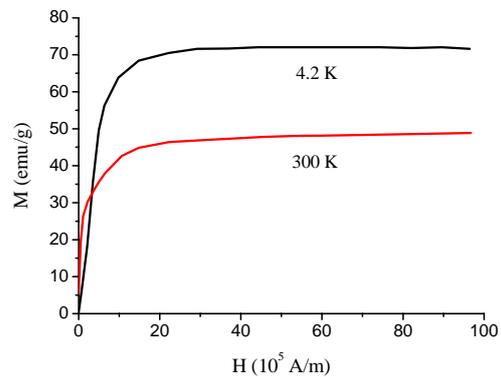

Fig. 3. Hysteresis loops of $BaFe_{12}O_{19}$ powder at 4.2 K (red line) and 300 K.

Fig. 4. Magnetization dependence on the magnetic field measured at 4.2 K and 300 K.

The magnetic measurements results show saturation magnetization $M_s$ of 48.86 emu/g at room temperature, which is lower than the theoretical one calculated for single crystals of barium hexaferrite particles, i.e. 72 emu/g, as reported by Shirk and Buessem [13]. This value is comparable to that of $BaFe_{12}O_{19}$ powder obtained by co-precipitation with particles size near the critical diameter [14]. This low saturation magnetization value can be explained by the fact that the particles are monodomain. Several theories, involving surface effects, spin canting and sample inhomogeneity, have been proposed to account for the relatively low magnetization in fine particles [15, 16]. The squareness ratio ($M_r/M_s$) is found to be around 0.5, which is close to the value expected for randomly packed single domain particles [17]. The coercivity value is 2.4 x $10^5$ A/m for both temperatures - 300 K and 4.2 K. The lower coercivity compared with that for a $BaFe_{12}O_{19}$ single crystal may be due to the fact that a part of the particles in the sample do not have a perfect hexagonal shape, so that structural defects exist, which may also cause the reduction in the sample coercivity. Kubo et al. [18] investigated the particles shape effect on the coercivity of hexaferrites. They found that $H_c$ decreased with increasing $D/t$. On the other hand, Chang et al. [19] showed that incoherent reversal was occurring in the single domain particles with diameter greater than 60 nm and the mode of reversal was extremely dependent on the thickness of the particles. Thus, it is believed that the lower coercivity may be caused not only by the larger aspect ratio, but also by an incoherent magnetization reversal.

Table 1  Magnetic properties of barium hexaferrite powder.

| T [K] | $M_s$ [emu/g] | $M_r$ [emu/g] | $M_r/M_s$ | $H_c$ $10^5$ [A/m] |
|---|---|---|---|---|
| 4.2 | 71.61 | 33.72 | 0.49 | 2.4 |
| 300 | 48.86 | 23.89 | 0.49 | 2.4 |

The law of approach to saturation was used to estimate the anisotropy field $H_a$ and the magneto-crystalline anisotropy $K_1$ [20]:

$$M = M_s(1 - \frac{A}{H} - \frac{B}{H^2}...) + \chi_p H \qquad (1)$$

where $A$ is the inhomogeneity parameter, $B$ is the anisotropy parameter and $\chi_p$, the high-field differential susceptibility. In the case of a hexagonal symmetry, $B$ may be expressed as [21]

$$B = \frac{H_a^2}{15} = \frac{4K_1^2}{15M_s^2} \qquad (2)$$

A linear relationship was found between $M$ and $1/H^2$ at magnetic fields ranging from 1.2 x $10^6$ to 2.3 x $10^6$ A/m. Thus, $H_a$ and $K_1$ can be calculated from Eq. 2. The results for $H_a$ and $K_1$ are 1.4 x $10^6$ A/m and 2.37 x $10^5$ J/m$^3$, respectively. The $K_1$ value is lower than that for bulk $BaFe_{12}O_{19}$.

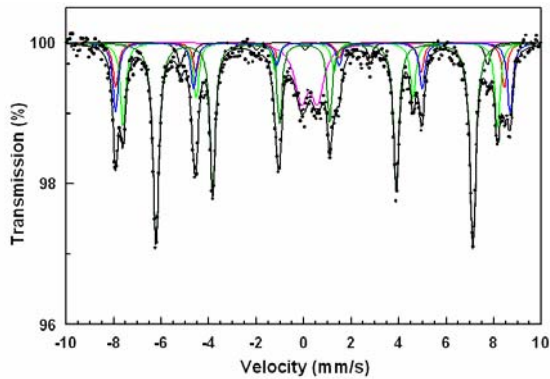

Fig. 5.  Mössbauer spectrum of $BaFe_{12}O_{19}$ powder at 300K.

Figure 5 shows a Mössbauer spectrum of $BaFe_{12}O_{19}$ powder at room temperature. The spectrum was fitted with five six-line sub-patterns and a doublet. The five six-line sub-patterns were assigned to the $12k$, $4f_2$, $4f_1$, $2a$ and $2b$ sites of the hexagonal crystal structure. It was assumed that the ratio of the areas for each six-line patterns was 3:2:1:1:2:3 and for the doublet, 1:1. The presence of a doublet in the Mössbauer spectrum may be due to the superparamagnetic behavior of the smallest particles. The Mössbauer spectroscopy results are summarized in Table 2. The sequence of hyperfine magnetic field ($H_{hf}$), isomer shift ($\delta_{Fe}$) and quadrupole shift ($2\varepsilon$) is the same as in the samples with mono-domain $BaFe_{12}O_{19}$ particles produced by co-precipitation methods [22]. The relative areas ($RA$) of Fe on the different sites do not correspond fully to the standard values (50 : 16.6 : 16.6 : 8.3 : 8.3 [23]). A similar distribution has been also observed in powder samples of mono-domain barium hexaferrite powders produced by co-precipitation [22, 24]. The bipyramidal ($2b$) sites are not fully occupied. These data demonstrate that as the hexagonal particles are formed and grow, the sites with high crystallographic symmetry are filled with priority. The increase in the $c$ cell parameter may be due to the incomplete occupation of the $2b$ sites by $Fe^{3+}$ ions. It is well known that the $2b$ site has the greatest contribution to the anisotropy of $BaFe_{12}O_{19}$. As was mentioned above, the technological conditions affect the cation distribution in barium hexaferrite, so that, as a result, the superexchange interaction related to the $Fe^{3+}$ distribution in the hexaferrite sublattices may be disturbed. Thus, the lower $K_1$ value calculated may be due to the incomplete filling of the $2b$ sites by $Fe^{3+}$ ions.

Table 2 Hyperfine parameters of barium hexaferrite powder produced by the microemulsion method.

|      | $H_{hf} 10^7$ [A/m] | $\delta_{Fe}$ [mm/s] | $2\varepsilon$ [mm/s] | RA [%] |
|------|---------------------|----------------------|------------------------|--------|
| $12k$  | 3.28                | 0.35                 | 0.42                   | 50     |
| $4f_2$ | 4.10                | 0.38                 | 0.20                   | 16     |
| $4f_1$ | 3.89                | 0.26                 | 0.24                   | 19     |
| $2a$   | 4.03                | 0.34                 | 0.06                   | 10     |
| $2b$   | 3.19                | 0.27                 | 2.23                   | 5      |

**Conclusion**

In summary, a microemulsion process for the synthesis of fine and uniform particles of barium hexaferrite is presented. This process allows us to obtain mono-domain $BaFe_{12}O_{19}$ particles, which are phase pure as confirmed by X-ray diffraction. The magnetic properties of this samples include saturation magnetization of 48.86 emu/g and 71.61 emu/g, respectively for 300 K and 4.2 K and a coercivity of $2.4 \times 10^5$ A/m for both temperatures, which may be attributed to the mono-domain structure of the particles and to the presence of a very small fraction of particles with size of about 100 nm. The lower value obtained for the magneto-crystalline anisotropy $K_1$ is most probably due to the incomplete filling of the bipyramidal $2b$ sites by $Fe^{3+}$ ions, as established by the analysis of the Mössbauer spectrum.

**Acknowledgements**

The work was supported in part by National Science Fund of Republic of Bulgaria under grants DO 02-99, DO 02-343 and DO 02-224, a research agreements between CGRI, Belgium and the Bulgarian Academy of Sciences, between F.W.O.-Flanders, Belgium and the Bulgarian Academy of Sciences, and a Joint Research Project between the Polish Academy of Sciences and the Bulgarian Academy of Sciences.